\documentclass[reqno,11pt]{amsart}
\usepackage[utf8]{inputenc}
\usepackage{enumitem}
\usepackage{graphicx}
\usepackage{amscd}
\usepackage{slashed}
\usepackage{amssymb}
\usepackage{mathtools} 
\usepackage{pstricks}
\usepackage[mathscr]{eucal}
\numberwithin{equation}{section}
\allowdisplaybreaks[1]

\title[Dark Matter and Energy in Causal Fermion Systems]{A Mechanism for Dark Matter and Dark Energy in the
Theory of Causal Fermion Systems}

\author[F.\ Finster]{Felix Finster}
\address{Fakult\"at f\"ur Mathematik \\ Universit\"at Regensburg \\ D-93040 Regensburg \\ Germany}%
\email{finster@ur.de}

\author[J.M.\ Isidro]{Jos{\'e} M. Isidro \\ \\ September 2022}
\address{Instituto Universitario de Matem\'atica Pura y Aplicada \\ Universidad Polit\'ecnica de Valencia \\
Valencia 46022 \\ Spain}
\email{joissan@mat.upv.es}

\newtheorem{Def}{Definition}[section]

\newtheorem{Prp}[Def]{Proposition}

\newtheorem{Remark}[Def]{Remark}

\newcommand{\Thanks}{\vspace*{.5em} \noindent \thanks}
\newcommand{\beq}{\begin{equation}}
\newcommand{\eeq}{\end{equation}}
\newcommand{\Proof}{\begin{proof}}
\newcommand{\QED}{\end{proof} \noindent}
\newcommand{\QEDrem}{\ \hfill $\Diamond$}

\newcommand{\la}{\langle}
\newcommand{\ra}{\rangle}

\newcommand{\Sl}{\mathopen{\prec}}
\newcommand{\Sr}{\mathclose{\succ}}

\newcommand{\C}{\mathbb{C}}
\newcommand{\R}{\mathbb{R}}
\newcommand{\1}{\mbox{\rm 1 \hspace{-1.05 em} 1}}
\newcommand{\Z}{\mathbb{Z}}
\newcommand{\N}{\mathbb{N}}
\newcommand{\Pdd}{\mbox{$\partial$ \hspace{-1.2 em} $/$}}

\renewcommand{\H}{\mathscr{H}}

\newcommand{\bep}{\begin{pmatrix}}
\newcommand{\enp}{\end{pmatrix}}

\renewcommand{\O}{\mathscr{O}}

\newcommand{\F}{{\mathscr{F}}}
\newcommand{\Dir}{{\mathcal{D}}}

\renewcommand{\O}{{\mathscr{O}}}
\renewcommand{\L}{{\mathcal{L}}}
\newcommand{\Sact}{{\mathcal{S}}}
\newcommand{\s}{{\mathfrak{s}}}

\newcommand{\Lin}{\text{\rm{L}}}

\newcommand{\T}{{\mathscr{T}}}

\newcommand{\Cisc}{C^\infty_{\text{\rm{sc}}}}

\newcommand{\scrM}{\myscr M}
\newcommand{\scrN}{\myscr N}

\DeclareFontFamily{OT1}{rsfso}{}
\DeclareFontShape{OT1}{rsfso}{m}{n}{ <-7> rsfso5 <7-10> rsfso7 <10-> rsfso10}{}
\DeclareMathAlphabet{\myscr}{OT1}{rsfso}{m}{n}

\DeclareMathOperator{\re}{Re}

\DeclareMathOperator{\Tr}{Tr}
\DeclareMathOperator{\tr}{tr}

\DeclareMathOperator{\supp}{supp}

\newcommand{\bitem}{\begin{itemize}[leftmargin=2.5em]}
\newcommand{\eitem}{\end{itemize}}

\begin{document}

\maketitle

\begin{abstract}
It is shown that the theory of causal fermion systems gives rise to a novel mechanism 
for dark matter and dark energy. This mechanism is first worked out for cubical subsets of Minkowski space
with periodic boundary conditions. Then it is studied in Friedmann-Lema{\^i}tre-Roberson-Walker spacetimes.
The magnitude of the effect scales like one over the lifetime of the universe squared.
In contrast to most models of dark matter and dark energy, our mechanism does not
postulate any new particles. Instead, it is a result of the collective behavior of all the wave functions which form the
Dirac sea, needed in order to arrange correlated initial and end quantum states of the universe.
\end{abstract}

\tableofcontents

\section{Introduction} \label{secintro}

The theory of {\em{causal fermion systems}} is a recent approach to fundamental physics
(see the basics in Section~\ref{secprelim}, the reviews~\cite{dice2014, nrstg, grossmann},
the textbooks~\cite{cfs, intro} or the website~\cite{cfsweblink}).
In this approach, spacetime and all objects therein are described by a measure~$\rho$
on a set~$\F$ of linear operators on a Hilbert space~$(\H, \la .|. \ra_\H)$. 
The physical equations are formulated by means of the so-called {\em{causal action principle}},
a nonlinear variational principle where an action~$\Sact$ is minimized under variations of the measure~$\rho$.

In the present paper we show that the theory of causal fermion systems gives rise to a novel
mechanism for dark matter and dark energy with the scaling behavior
\begin{align}
\text{dark energy:} \qquad & \Lambda \sim \frac{1}{T^2} \label{de} \\
\text{dark matter:} \qquad & T^0_0 \sim \frac{1}{G \,T^2} \:, \label{dm}
\end{align}
where~$G$ is the gravitational constant, and~$T$ is the lifetime of the universe.
Here by ``lifetime of the universe'' we mean the time from
the ``big bang'' until reaching the final state of the universe.
For this ``final state'' there are various scenarios: big crunch, big bounce, big rip or the heat death
(for an overview see the textbook~\cite{ellis-mccallum}).
The physical value of~$T$ is unknown and depends on the specific cosmological model
(for a few references on this question see~\cite{donpage, bousso-freivogel}).
In order to get a first idea of the scalings,
one can choose~$T$ as the current age of the universe,
\beq \label{Texperiment}
T \approx 14 \text{ billion light years} \:.
\eeq
In this case, one obtains in Planck units
\[ \frac{1}{T^2} \approx 1.6 \times 10^{-122} \:, \]
reproducing the correct orders of magnitude
of the experimental values for dark energy and dark matter.
This means that, at least for cosmological scenarios in which the lifetime of the universe
has the same order of magnitude as the present age of the universe, our mechanism gives
a possible explanation for the appearance of dark matter and dark energy, including the correct
order of magnitude of its experimental values.
In particular, one gets a simple understanding of why the cosmological constant is so small.

In order to put our results into context, we begin with a few general
words on dark matter and dark energy (for a general account see~\cite{penrosebook}).
There are an estimated 10 billion galaxies (1~billion $=10^9$) in the observable portion of the universe. Most of these galaxies are in recessional motion away from each other according to Hubble's linear law \cite{weinberg-cosmology}.  Moreover, the farthest-lying galaxies appear to recede from us at speeds higher than predicted by Hubble's simple law. Indeed very precise experiments \cite{PERLMUTTER, RIESS} have shown that this recession is accelerating exponentially in cosmic time so that, at large distances, gravity appears to develop a repulsive character. There appears to be an energy component acting as an accelerating agent in the Cosmos; this energy component has been called {\it dark energy}\/. One possible explanation for it is a non-vanishing cosmological constant $\Lambda$ in Einstein's field equations; current data~\cite{planck2018} suggest the tiny value $\Lambda=2.9\times 10^{-122}$ in natural units. For a detailed exposition of the cosmological constant and related theories the reader may for example consult the references~\cite{CARROLL, weinberg-cc, singde, padmanabhande, ashtekarcc, faraggi}. Ultimately, any viable theory of quantum gravity (see for example~\cite{kiefernew}) must be
in a position to explain these conundrums. A first ansatz towards this goal is provided by the
holographic principle~\cite{thooft, susskind, bousso}.

Moreover, there is a large fraction of unknown material within the observed universe that has been dubbed {\it dark matter} (for general references see~\cite{bahcall, milgrom}). This dark matter is non-luminous and interacts with baryonic matter essentially via the gravitational interaction only. The  amount of evidence supporting this conclusion is impressive; for details see for
example~\cite{TRIMBLE}. Here we will just mention a few key pieces of data  suggesting that there is more matter in the universe than meets the eye: the rotation curves of galaxies; velocities in galaxy clusters; X-ray emission by galaxy clusters; gravitational lensing by galaxy clusters. For example, the rotation velocities of stars around a center of attraction do not follow the usual Kepler law; instead they exhibit a constant rotational velocity, even when the stars are very far out from the center of the galaxy. All this implies that there is an enormous amount of matter of non-baryonic origin still to be accounted for. 

Altogether, our universe is approximately flat, its mass/energy budget being roughly given by  $b=4.9\%$ (baryonic matter), $d_m=26.8\%$ (dark matter) and $d_e=68.3\%$ (dark energy). Thus our knowledge of matter within the universe hardly accounts for $5\%$ of its total, which makes the challenge to understand the remaining $95\%$ all the more pressing.

Numerous theories of the dark universe have been put forward, too many to be quoted here. 
In the present paper we propose a novel explanation for the dark universe (both dark matter and dark energy).
Our mechanism is based on the specific structures of a causal fermion system and the 
specific form of the causal action principle.
The causal action principle can be regarded as a variational principle for a family of
spinorial wave functions in spacetime (for a basic introduction to causal fermion
systems and the causal action principle see the preliminaries in Section~\ref{secprelim}).
In the limiting case of Dirac wave functions, this family includes all the solutions of
negative energy which form the Dirac sea. The resulting Dirac sea configuration is a minimizer
of the causal action principle. This entails that the states forming the Dirac sea are ``invisible''
in the sense that they do not enter the Euler-Lagrange (EL) equations of the causal action principle.
In the so-called continuum limit as worked out in~\cite{cfs}, the EL equations of the causal action principle
give rise to classical field equations (like the Maxwell, Yang-Mills and Einstein equations).
In this way, the causal action principle gives an explanation for why the Dirac sea does not enter
the classical field equations (in particular, the naively obtained ``infinite negative energy density'' of the
Dirac sea drops out).
However, {\em{deviations from the vacuum Dirac sea structure}} do {\em{not}} drop out of the field equations.
This fact is crucial for the derivation of the Einstein equation from the causal action principle,
because the deviation of the Dirac sea structure due to the curvature of spacetime gives rise to
the Ricci tensor in the Einstein equations (for details see~\cite[Sections~4.5 and~4.9]{cfs}).

Our mechanism for dark matter and dark energy can also be understood along these lines,
because it is a consequence of a deviation from the Dirac sea structure.
But now the deviation is a consequence of the fact that we demand the physical system
to be in a correlated quantum state both at its beginning (``big bang'') and at its end
(big crunch, big rip, big bounce or thermal death). In simple terms, this means that all the wave functions
forming the Dirac sea should be ``in phase'' at both the initial and the final state.
The resulting boundary conditions make it necessary to slightly modify each wave function.
The collective effect of all these small adjustments of Dirac wave functions
amounts to a modification of the Dirac sea configuration which
shows up in the Einstein equations as
dark matter and dark energy with the scalings~\eqref{de} and~\eqref{dm}.

Before explaining this mechanism in some more detail, we point out that for
our mechanism we do not need to postulate any new
particles (like axions, WIMPs or supersymmetric partners).
In fact, what appears as dark energy and dark matter is not to be considered
as ``matter'' or ``energy'' in the usual sense.
In this way, our approach bears some similarity with proposals made in~\cite{GEORGI, verlinde17}. However, in contrast to these proposals,
our mechanism is a result of a collective behavior of all the wave functions of the
Dirac sea, needed in order to obtain correlated initial and end states of the universe.

We conclude the introduction by describing
our mechanism in some more detail in the simplest possible setting:
a {\em{time-periodic, flat universe}} obtained by identifying~$t=0$ with~$t=T$ in Minkowski space.
Our starting point is the measure~$\rho$ describing the vacuum in Minkowski space
(for details see Section~\ref{secmink}).
The resulting causal fermion system can be characterized equivalently by the family of
physical wave functions, which can be understood in simple terms as describing the ``occupied one-particle states''
of the system. An important feature of a causal fermion system is that the physical wave functions
involve a regularization.
We assume that a suitably regularized Dirac sea configuration is a minimizer of the causal action principle.
For the purpose of the present paper, we do not need to be specific about the detailed form
of the regularization (see also the discussion of this point in Section~\ref{secmink}).
For this reason, it suffices to consider the simplest possible regularization by a
convergence-generating exponential factor. The resulting family of regularized
physical wave functions is described by the {\em{regularized kernel of the fermionic projector}}
\beq \label{Peps}
P^\varepsilon(x,y) = \int_{\R^4} \frac{d^4p}{(2 \pi)^4} \: (\slashed{p}+m)\:
\delta(p^2 - m^2)\: \Theta(-p^0)\: e^{\varepsilon p^0} e^{-ip (x-y)}
\eeq
(where~$p(x-y)$ denotes the Minkowski inner product and~$\varepsilon>0$ is the regularization length).

Taking this kernel as the starting point, we analyze effects which arise
because spacetime has a finite lifetime. As mentioned before, by
``lifetime of the universe'' we mean the time from
the ``big bang'' until the final state of the universe is reached.
Here, for technical simplicity, we consider the simplest possible realization of a finite lifetime,
namely a {\em{time-periodic}} universe.
In preparation, we consider the effects
of {\em{finite spatial volume}}. Indeed, as we shall see later in this paper, finite spatial volume will have no
measurable consequences, and for the analysis and results of the present paper it makes no difference
whether the spatial volume is finite or infinite. Nevertheless, considering a system in
finite spatial volume is useful for introducing our concepts.
Suppose we consider a system in a three-dimensional box of length~$2L$. Thus we replace
Minkowski space by the spacetime cylinder
\[ \scrM = \R \times [-L,L]^3 \]
Choosing periodic boundary conditions, this means that the spatial momenta~$\vec{k}$ must be
on a three-dimensional lattice of spacing~$L/\pi$, i.e.\
\[ \vec{k} \in {\mathfrak{L}} := \Big( \frac{\pi}{L} \:\Z \Big)^3 \:. \]
Likewise, in~\eqref{Peps} one replaces the integral over the momenta by a sum,
\[ P^\varepsilon_L(x,y) := \frac{1}{(2 \pi)^4}
\int_{-\infty}^\infty d\omega \: \Big( \frac{\pi}{L} \Big)^3 \sum_{\vec{k} \in {\mathfrak{L}}}
\: (\slashed{p}+m)\:
\delta(p^2 - m^2)\: \Theta(-p^0)\: e^{\varepsilon \omega} e^{-ip (x-y)} \:, \]
where~$p :=(\omega, \vec{k})$. These discrete momenta can also be realized by
summing over the dual lattice in position space, i.e.
\[ P^\varepsilon_L(x,y) = \sum_{\vec{\ell} \in (2 L \Z)^3}
P^\varepsilon\big( x,y + (0,\vec{\ell})\big) \:. \]
By summing over the lattice, we arrange the periodic boundary conditions, because, shifting
the summation index,
\begin{align}
P^\varepsilon_L \big( x,y +(0, \vec{\ell'}) \big) &= \sum_{\vec{\ell} \in (2 L \Z)^3}
P^\varepsilon\big( x,y + (0,\vec{\ell}+\vec{\ell'})\big) \notag \\
&= \sum_{\vec{\ell} \in (2 L \Z)^3}
P^\varepsilon\big( x,y + (0,\vec{\ell})\big) = P^\varepsilon_L (x,y) \:. \label{periodspace}
\end{align}
We refer to this procedure as {\em{periodization}}.
Since~$P^\varepsilon$ decays exponentially for large spatial distances like~$\sim \exp(-m\, |\vec{x}-\vec{y}|)$
(for details see Section~\ref{secperiodize} below),
spatial periodization gives rise to contributions which decay exponentially
in the size~$L$ of the box; this is why spatial periodization does not give rise to any
measurable physical effects.

\begin{figure}[tb]
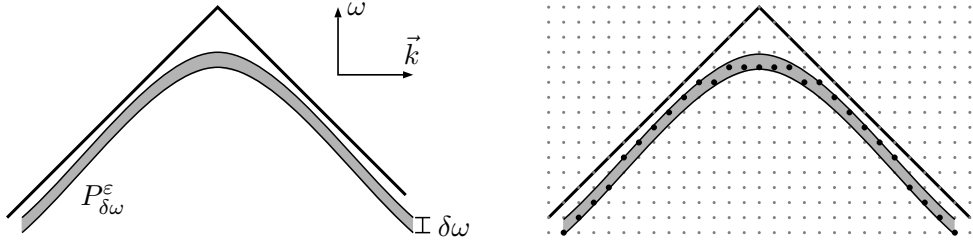

\psscalebox{1.0 1.0} 
{

}
\caption{Periodizing a smeared-out Dirac sea configuration.}
\label{figperiodize}
\end{figure}

The situation becomes more interesting if the system should also be {\em{periodic in time}}
with period~$T$ (which is to be thought of as the total life time of the universe; for practical purposes
we can only take~$T$ as the current age of the universe as given and discussed in~\eqref{Texperiment}).
A simple periodization in time does {\em{not}} work, because the series
\beq \label{periodtnaive}
\sum_{\tau \in T \Z} P^\varepsilon_L\big( x,y + (\tau,0)\big)
\eeq
diverges. One way of understanding how this comes about is to note that~$P(x,y)$ decays
for large times only polynomially, at a rate which is not fast enough for the series to converge
(for details see again Section~\ref{secperiodize} below).
Alternatively, the problem of~\eqref{periodtnaive} can be understood from the
fact that making the system also periodic in time means that the momenta~$p$ must be
on the four-dimensional lattice
\[ p \in \frac{2 \pi}{T}\:\Z \;\times\; \Big( \frac{\pi}{L} \:\Z \Big)^3 \:. \]
However, for generic values of~$m$, this lattice does not intersect the mass hyperbola~$p^2=m^2$.
In order to improve the situation, one needs to {\em{smear out}} the Dirac sea configuration,
for example by replacing the $\delta$-distribution in~\eqref{Peps} by a multiple times a
characteristic function supported in a ``strip of width~$\delta \omega$'' around the mass shell
(see the left of Figure~\ref{figperiodize}; the detailed form of this smearing will be
introduced below).
Denoting the resulting distribution by~$P^\varepsilon_{\delta \omega}$,
periodizing in space and time becomes well-defined,
\beq \label{periodized}
P^\varepsilon_{\delta \omega, T, L}(x,y)
:= \sum_{\tau \in T \Z} \;\sum_{\vec{\ell} \in (2 L \Z)^3}
P^\varepsilon_{\delta \omega} \big( x,y + (\tau,\vec{\ell})\big) \:.
\eeq
It corresponds in momentum space
to a multiplication by $\delta$-distributions supported on a four-dimensional lattice
(see the right of Figure~\ref{figperiodize}).

Before discussing the role of the smearing in detail, we point out that the above construction
leads to a violation of the Dirac equation, because the mass parameter is no longer fixed,
but has been ``smeared out.'' In other words, the time periodization makes it necessary
to go beyond the standard physical description. In the theory of causal fermion systems, where the Dirac
and Einstein equations emerge only in a limiting case (the so-called continuum limit as worked out
in~\cite{cfs}), modifications of the Dirac equation arise naturally and do not cause any conceptual
difficulties. One should also note that the energy scale of the smearing is of the order
\[ \delta \omega \sim \frac{1}{T} \:. \]
Compared to the Compton scale, this energy scale is extremely small. Therefore, the smearing
is a tiny effect. Nevertheless, as we shall work out in detail in what follows,
it will lead to corrections to the Einstein equations which can be interpreted as
dark energy and dark matter with the scalings~\eqref{de} and~\eqref{dm}.

We next explain how the smearing is performed. Following the above assumption that
the regularized kernel~$P^\varepsilon(x,y)$ in Minkowski space must
correspond to a minimizer of the causal action principle,
our guiding principle is that the smearing and periodization in time should not leave the
causal Lagrangian as far as possible unchanged. Moreover, we want to preserve
\bitem
\item[{\rm{(a)}}] the number of occupied states
\item[{\rm{(b)}}] the normalization of the states
\eitem
These two requirements can be understood from the fact that the corresponding objects and quantities
have a direct relation to inherent structures of the causal fermion system. Indeed, the number of occupied states
corresponds to the
dimension of the Hilbert space~$\H$, whereas the normalization of the states 
and the corresponding unitary time evolution are related 
to a conservation law resulting from a Noether-like theorem for the causal action principle
(more precisely, the conservation law for the commutator inner product; for details see~\cite{noether, dirac}).
Keeping the number of occupied states fixed leads us to choose the width~$\delta \omega$
independent of~$\vec{k}$ as
\beq \label{delomegaintro}
\delta \omega = \frac{2 \pi}{T} \:,
\eeq
because with this choice, for every~$\vec{k} \in {\mathfrak{L}}$ there is exactly one
four-dimensional lattice point inside the strip (see the right of Figure~\ref{figperiodize}).
In order to keep the normalization unchanged, we choose the weight function inside the strip
to be constant. Moreover, we describe the states inside the strip again by solutions of the Dirac equation,
but with a ``smeared'' mass. Apart from being conceptually simple, this procedure also has the
advantage of carrying over to curved spacetime (as will be explained
in Section~\ref{secflrw} in the FLRW spacetime). This leads us to the ansatz
\beq \label{Psmearintro}
\hat{P}^\varepsilon_{\delta \omega}(p)
= \frac{1}{\delta m(\vec{k})} \int_{m_0+\delta n(\vec{k})}^{m_0+\delta m(\vec{k}) + \delta n(\vec{k})} 
(\slashed{p}+m)\:
\delta(p^2 - m^2)\: \Theta(-p^0)\: e^{\varepsilon p^0}\: dm \:,
\eeq
where~$m_0$ is the unsmeared mass, and~$\delta m(\vec{k})$ is chosen in such a way that the corresponding
frequency shift~$\delta \omega$ as given by
\beq \label{disperse}
\delta \omega = \sqrt{|\vec{k}|^2 + \big(m+\delta m(\vec{k}) \big)^2} - 
\sqrt{|\vec{k}|^2 + m^2}
\eeq
satisfies~\eqref{delomegaintro} (for details see the computation after~\eqref{delomega}).
The function~$\delta n(\vec{k})$ gives us the freedom to slightly modify
the shape of the mass hyperbola with the goal of compensating the leading
effect of the smearing (for details see Remark~\ref{remn} below).
Working out all the relevant effects in detail, we find contributions to the Einstein equations
of the form~\eqref{de} and~\eqref{dm}.

The paper is organized as follows. After the necessary preliminaries (Section~\ref{secprelim})
we introduce causal fermion systems in Minkowski space (Section~\ref{secmink}).
In Section~\ref{secdarkmatter} we compute the effect of the smearing and periodization in detail
and derive~\eqref{de} and~\eqref{dm}. In Section~\ref{secflrw} it is explained how the methods and results carry over to a Friedmann-Lema{\^i}tre-Robertson-Walker spacetime.
In Section~\ref{secoutlook} we conclude the paper with a brief discussion and outlook.

\section{Preliminaries} \label{secprelim}
This section provides the necessary background on causal fermion systems.

\subsection{Causal Fermion Systems and the Causal Action Principle}
We begin with the general definitions.
\begin{Def} \label{defcfs} (causal fermion systems) {\em{ 
Given a separable complex Hilbert space~$\H$ with scalar product~$\la .|. \ra_\H$
and a parameter~$n \in \N$ (the {\em{spin dimension}}), we let~$\F \subset \Lin(\H)$ be the set of all
symmetric\footnote{Here by a (bounded) symmetric operator~$A$ we mean that~$\la A u | v \ra_\H =
\la u | A v \ra_\H$ for all~$u,v \in \H$. Representing the operator in
an orthonormal basis, the resulting matrix is Hermitian.
For bounded operators as considered here,
the notions ``symmetric'' and ``self-adjoint'' coincide.} 
 operators on~$\H$ of finite rank, which (counting multiplicities) have
at most~$n$ positive and at most~$n$ negative eigenvalues. On~$\F$ we are given
a positive measure~$\rho$ (defined on a $\sigma$-algebra of subsets of~$\F$).
We refer to~$(\H, \F, \rho)$ as a {\em{causal fermion system}}.
}}
\end{Def} \noindent
A causal fermion system describes a spacetime together
with all structures and objects therein.
The physical equations are formulated for a causal fermion system
by demanding that the measure~$\rho$ should be a minimizer of the causal action principle
defined as follows. For any~$x, y \in \F$, the product~$x y$ is an operator of rank at most~$2n$. 
However, in general it is no longer symmetric because~$(xy)^* = yx$,
and this is different from~$xy$ unless~$x$ and~$y$ commute.
As a consequence, the eigenvalues of the operator~$xy$ are in general complex.
We denote the non-trivial eigenvalues counting algebraic multiplicities
by~$\lambda^{xy}_1, \ldots, \lambda^{xy}_{2n} \in \C$
(more specifically,
denoting the rank of~$xy$ by~$k \leq 2n$, we choose~$\lambda^{xy}_1, \ldots, \lambda^{xy}_{k}$ as all
the non-zero eigenvalues and set~$\lambda^{xy}_{k+1}, \ldots, \lambda^{xy}_{2n}=0$).
We introduce the Lagrangian and the causal action by
\begin{align}
\text{\em{Lagrangian:}} && \L(x,y) &= \frac{1}{4n} \sum_{i,j=1}^{2n} \Big( \big|\lambda^{xy}_i \big|
- \big|\lambda^{xy}_j \big| \Big)^2 \label{Lagrange} \\
\text{\em{causal action:}} && \Sact(\rho) &= \iint_{\F \times \F} \L(x,y)\: d\rho(x)\, d\rho(y) \:. \label{Sdef}
\end{align}
The {\em{causal action principle}} is to minimize~$\Sact$ by varying the measure~$\rho$
under the following constraints,
\begin{align}
\text{\em{volume constraint:}} && \rho(\F) = \text{const} \quad\;\; & \label{volconstraint} \\
\text{\em{trace constraint:}} && \int_\F \tr(x)\: d\rho(x) = \text{const}& \label{trconstraint} \\
\text{\em{boundedness constraint:}} && \iint_{\F \times \F} 
|xy|^2
\: d\rho(x)\, d\rho(y) &\leq C \:, \label{Tdef}
\end{align}
where~$C$ is a given parameter, $\tr$ denotes the trace of a linear operator on~$\H$, and
the absolute value of~$xy$ is the so-called spectral weight,
\[ 
|xy| := \sum_{j=1}^{2n} \big|\lambda^{xy}_j \big| \:. \]
This variational principle is mathematically well-posed if~$\H$ is finite-dimensional
(for details see~\cite{continuum} or~\cite[Chapter~12]{intro}).

A minimizer of the causal action principle
satisfies the following {\em{Euler-Lagrange (EL) equations}}.
For a suitable value of the parameter~$\s>0$,
the function~$\ell : \F \rightarrow \R_0^+$ defined by
\beq \label{elldef}
\ell(x) := \int_M \L_\kappa(x,y)\: d\rho(y) - \s
\eeq
is minimal and vanishes on the support\footnote{The
{\em{support}} of a measure is defined as the complement of the largest open set of measure zero, i.e.
\[ \supp \rho := \F \setminus \bigcup \big\{ \text{$\Omega \subset \F$ \,\big|\,
$\Omega$ is open and $\rho(\Omega)=0$} \big\} \:. \]
It is by definition a closed set.}
of~$\rho$,
\beq \label{EL}
\ell|_{\supp \rho} \equiv \inf_\F \ell = 0 \:.
\eeq
Here the $\kappa$-{\em{Lagrangian}}~$\L_\kappa$ is defined by
\beq \label{Lkappa}
\L_\kappa \::\: \F \times \F \rightarrow \R\:,\qquad
\L_\kappa(x,y) := \L(x,y) + \kappa\: |xy|^2
\eeq
with a non-negative parameter~$\kappa$, which can be thought of as the Lagrange parameter
corresponding to the boundedness constraint.
Likewise, the parameter~$\s \geq 0$ in~\eqref{elldef} is the Lagrange parameter
corresponding to the volume constraint. For the derivation and further details we refer to~\cite[Section~2]{jet}
or~\cite[Chapter~7]{intro}.

\subsection{Spacetime and Causal Structure}
Let~$\rho$ be a {\em{minimizing}} measure. {\em{Spacetime}}
is defined as the support of this measure,
\[ 
M := \supp \rho \;\subset\; \F \:, \]
where on~$M$ we consider the topology induced by~$\F$ (generated by the operator norm
on~$\Lin(\H)$).
Thus the spacetime points are symmetric linear operators on~$\H$.
The restriction of the measure~$\rho|_M$ gives a volume measure on spacetime.

The operators in~$M$ contain a lot of information which, if interpreted correctly,
gives rise to spacetime structures like causal and metric structures, spinors
and interacting fields (for details see~\cite[Chapter~1]{cfs}).
All the resulting objects are {\em{inherent}} in the sense that we only use information
already encoded in the causal fermion system.
Here we restrict attention to those structures needed in what follows.
We begin with the following notion of causality:

\begin{Def} (causal structure) \label{def2} 
{\em{ For any~$x, y \in \F$, we again denote the non-trivial ei\-gen\-values of the operator product~$xy$
(again counting algebraic multiplicities) by~$\lambda^{xy}_1, \ldots, \lambda^{xy}_{2n}$.
The points~$x$ and~$y$ are
called {\em{spacelike}} separated if all the~$\lambda^{xy}_j$ have the same absolute value.
They are said to be {\em{timelike}} separated if the~$\lambda^{xy}_j$ are all real and do not all 
have the same absolute value.
In all other cases (i.e.\ if the~$\lambda^{xy}_j$ are not all real and do not all 
have the same absolute value),
the points~$x$ and~$y$ are said to be {\em{lightlike}} separated. }}
\end{Def} \noindent
Restricting the causal structure of~$\F$ to~$M$, we get causal relations in spacetime.

The Lagrangian~\eqref{Lagrange} is compatible with the above notion of causality in the
following sense.
Suppose that two points~$x, y \in M$ are spacelike separated.
Then the eigenvalues~$\lambda^{xy}_i$ all have the same absolute value.
As a consequence, the Lagrangian~\eqref{Lagrange} vanishes. Thus pairs of points with spacelike
separation do not enter the action. This can be seen in analogy to the usual notion of causality where
points with spacelike separation cannot influence each other.
This is the reason for the notion ``causal'' in {\em{causal}} fermion system
and {\em{causal}} action principle.

\subsection{Spinors and Physical Wave Functions} \label{secinherent}
A causal fermion system also gives rise to spinorial wave functions in spacetime, as we now explain.
For every~$x \in \F$ we define the {\em{spin space}}~$S_x$ by~$S_x = x(\H)$;
it is a subspace of~$\H$ of dimension at most~$2n$.
It is endowed with the {\em{spin inner product}} $\Sl .|. \Sr_x$ defined by
\beq \label{ssp}
\Sl u | v \Sr_x = -\la u | x v \ra_\H \qquad \text{(for all $u,v \in S_x$)} \:.
\eeq
It is an important observation that every vector~$u \in \H$ of the Hilbert space gives rise to a unique
wave function denoted by~$\psi^u$, which to every~$x \in M$ associates a vector of the
corresponding spin space~$\psi^u(x) \in S_x$. It is obtained by orthogonal projection to the
spin space,
\beq \label{psiudef}
\psi^u \::\: M \rightarrow \H\qquad \text{with} \qquad \psi^u(x) := \pi_x u \in S_xM \quad \text{for all~$x \in M$}\:.
\eeq
We refer to~$\psi^u$ as the {\em{physical wave function}} of the vector~$u \in \H$.
Varying the vector~$u \in \H$, we obtain a whole family of physical wave functions.
This family is described most conveniently by the
{\em{wave evaluation operator}}~$\Psi$ defined at every spacetime point~$x \in M$ by
\beq \label{weo}
\Psi(x) \::\: \H \rightarrow S_x \:, \qquad u \mapsto \psi^u(x) \:.
\eeq
It is a simple but important observation that every spacetime point operator can be recovered from
its wave evaluation operator by (for the proof see for example~\cite[Lemma~1.1.3]{cfs}).
\beq \label{xPsi}
x = - \Psi(x)^* \Psi(x) \:.
\eeq
Having constructed the spacetime point operators, we also recover all the other
inherent structures of a causal fermion system. Proceeding in this way, all spacetime structures
can be regarded as being induced by the physical wave functions.
Moreover,  restricting attention to variations of~$\Psi$, one can understand the causal action principle as
a variational principle for the family of physical wave functions.
Finally, one can construct concrete examples
of causal fermion systems by choosing the physical wave functions more specifically as
the quantum mechanical wave functions
in a classical Lorentzian spacetime. In the next section we explain this construction in more detail.

\subsection{Describing a Lorentzian Spacetime by a Causal Fermion System} \label{seccfslorentz}
We now explain how a classical curved spacetime is described by a causal fermion system.
Our starting point is Lorentzian spin geometry.
Thus we let~$(\scrM, g)$ be a smooth, globally hyperbolic, time-oriented
Lorentzian spin manifold of dimension four.
For the signature of the metric we use the convention~$(+ ,-, -, -)$.
We denote the corresponding spinor bundle by~$S\scrM$. Its fibers~$S_x\scrM$ (with~$x \in \scrM$)
are endowed with an inner product~$\Sl .|. \Sr_x$ of signature~$(2,2)$.
Clifford multiplication is described by a mapping~$\gamma$
which satisfies the anti-commutation relations,
\[ 
\gamma \::\: T_x\scrM \rightarrow \Lin(S_x\scrM) \qquad
\text{with} \qquad \gamma(u) \,\gamma(v) + \gamma(v) \,\gamma(u) = 2 \, g(u,v)\,\1_{S_x(\scrM)} \:. \]
We also write Clifford multiplication in components with the Dirac matrices~$\gamma^j$.
The metric connections on the tangent bundle and the spinor bundle are denoted by~$\nabla$.

We denote the smooth sections of the spinor bundle by~$C^\infty(\scrM, S\scrM)$.
The Dirac operator~$\Dir$ is defined by
\[ \Dir := i \gamma^j \nabla_j \::\: C^\infty(\scrM, S\scrM) \rightarrow C^\infty(\scrM, S\scrM)\:. \]
Given a real parameter~$m \in \R$ (the {\em{mass}}), the Dirac equation reads
\[ (\Dir - m) \,\psi = 0 \:. \]
We mainly consider solutions in the class~$\Cisc(\scrM, S\scrM)$ of smooth sections
with spatially compact support (i.e.\ wave functions whose restriction to any Cauchy surface
is compact). On such solutions, one has the scalar product
\beq \label{print}
(\psi | \phi)_m = \int_\scrN \Sl \psi \,|\, \gamma(\nu)\, \phi \Sr_x\: d\mu_\scrN(x) \:,
\eeq
where~$\scrN$ denotes any Cauchy surface and~$\nu$ its future-directed normal
(due to current conservation, the scalar product is
in fact independent of the choice of~$\scrN$; for details see~\cite[Section~2]{finite}).
Forming the completion gives the Hilbert space~$(\H_m, (.|.)_m)$.

We next choose a closed subspace~$\H \subset \H_m$
of the solution space of the Dirac equation.
The induced scalar product on~$\H$ is denoted by~$\la .|. \ra_\H$.
There is the technical difficulty that the wave functions in~$\H$ are in general not continuous,
making it impossible to evaluate them pointwise.
For this reason, we need to introduce a {\em{regularization}} on
the length scale~$\varepsilon$, described mathematically by a linear
\beq \label{Reps}
\text{\em{regularization operator}} \qquad {\mathfrak{R}}_\varepsilon \::\: \H \rightarrow C^0(\scrM, S\scrM) \:.
\eeq
In the simplest case, the regularization can be realized by a convolution
on a Cauchy surface or in spacetime (for details see~\cite[Section~4]{finite}
or~\cite[Section~\S1.1.2]{cfs}). For us, the regularization is not merely a technical tool,
but it realizes the concept that we want to change the geometric structures on the microscopic
scale. With this in mind, we always consider the regularized quantities as those having mathematical and
physical significance. Different choices of regularization operators realize different
microscopic spacetime structures.

Evaluating the regularization operator at a spacetime point~$x \in \scrM$ gives 
the {\em{regularized wave evaluation operator}}~$\Psi^\varepsilon(x)$,
\beq \label{Pepsdirac}
\Psi^\varepsilon(x) = {\mathfrak{R}}_\varepsilon(x) \::\: \H \rightarrow S_x \scrM \:.
\eeq
We also take its adjoint (with respect to the Hilbert space
scalar product~$\la .|. \ra_\H$ and the spin inner product~$\Sl .|. \Sr_x$),
\[ \big(\Psi^\varepsilon(x) \big)^* \::\: S_x \scrM \rightarrow \H  \:. \]
Multiplying~$\Psi^\varepsilon(x)$ by its adjoint gives the operator
\beq \label{Fepsprod}
F^\varepsilon(x) := - \big( \Psi^\varepsilon(x) \big)^* \,\Psi^\varepsilon(x) \::\: \H \rightarrow \H \:,
\eeq
referred to as the {\em{local correlation operator}} at
the spacetime point~$x$. The local correlation operator is also characterized by the relation
\beq \label{Fepsdef}
(\psi \,|\, F^\varepsilon(x)\, \phi) = -\Sl ({\mathfrak{R}}_\varepsilon\psi)(x) | 
({\mathfrak{R}}_\varepsilon \phi)(x) \Sr_x \qquad \text{for all~$\psi, \phi
\in \H$} \:.
\eeq
Taking into account that the inner product on the Dirac spinors at~$x$ has signature~$(2,2)$,
it is a symmetric operator on~$\H$
of rank at most four, which (counting multiplicities) has at most two positive and at most two negative eigenvalues.
Varying the spacetime point, we obtain a mapping
\[ F^\varepsilon \::\: \scrM \rightarrow \F \subset \Lin(\H)\:, \]
where~$\F$ denotes all symmetric operators of
rank at most four with at most two positive and at most two negative eigenvalues.
Finally, we introduce the measure~$\rho$ on~$\F$ by taking
the push-forward of the volume measure on~$\scrM$ under the mapping~$F^\varepsilon$,
\beq \label{rhoeps}
\rho := (F^\varepsilon)_* \mu_\scrM
\eeq
(thus~$\rho(\Omega) := \mu_\scrM((F^\varepsilon)^{-1}(\Omega))$).
The resulting structure~$(\H, \F, \rho)$ is a causal fermion system of spin dimension two.

\subsection{Classical Gravity and Beyond} \label{seccl}
The construction of the causal fermion system in the previous section involves the
Lorentzian metric~$g$. But this Lorentzian metric does not need to satisfy the
Einstein equation. Instead of postulating the Einstein equations, one can {\em{derive}}
these equations from the causal action principle. To this end, one evaluates the
EL equations~\eqref{EL} for the causal fermion system~$(\H, \F, \rho)$ constructed in the
previous section in the limit~$\varepsilon \searrow 0$ when the
regularization is removed. This analysis, referred to as the {\em{continuum limit}}, is carried out
in detail in~\cite[Chapter~4]{cfs}. The main result is that the EL equations are satisfied
asymptotically for small~$\varepsilon>0$ only if the Lorentzian metric satisfies the Einstein equations,
up to possible higher order corrections in curvature (which scale in powers of~$(G\:
\text{Riem})$, where~$G$ is the gravitational coupling constant and~$\text{Riem}$ is the
curvature tensor), i.e.\ (see~\cite[Theorems~4.9.3 and~5.4.4]{cfs})
\beq \label{einstein}
R_{jk} - \frac{1}{2}\:R\: g_{jk} + \Lambda\, g_{jk} = G\, T_{jk} 
+ {\mathscr{O}} \big( G^2 \,\text{Riem}^2 \big) \:.
\eeq

We now briefly outline how this result is derived in~\cite[Chapter~4]{cfs}.
Given a causal fermion system~$(\H, \F, \rho)$ describing a globally hyperbolic spacetime~$(\scrM, g)$
(as constructed in Section~\ref{seccfslorentz}), the task is to evaluate the EL equations~\eqref{EL}.
For the computations, it is very useful that the EL equations can be expressed
in terms of the {\em{kernel of the fermionic projector}}~$P(x,y)$.
Here we do not need to enter the details
(which can be found in~\cite[\S1.1.3 and~\S1.4.1]{cfs}), but it suffices
to explain what the kernel of the fermionic projector is and how it encodes the gravitational field.
Abstractly, the kernel of the fermionic projector is defined in terms of the wave evaluation operator~\eqref{weo} by
\[ P(x,y) = -\Psi(x)\, \Psi(y)^* \::\: S_y \rightarrow S_x \:. \]
In an orthonormal basis~$(e_i)$ of~$\H$, it can be expressed as
\beq \label{Psum}
P(x,y) = -\sum_i |\psi^{e_i}(x) \Sr \Sl \psi^{e_i}(y)| \:,
\eeq
showing that it is composed of all the physical wave functions of the system
(for details see also~\cite[\S1.2.4]{cfs}).
For a causal fermion system constructed in a Lorentzian spacetime, one can identify~$P(x,y)$
with the {\em{regularized kernel of the fermionic projector}}~$P^\varepsilon(x,y)$ defined by
(for details see~\cite[Section~1.2]{cfs})
\beq \label{Pepsdef}
P^\varepsilon(x,y) = -\Psi^\varepsilon(x)\big( \Psi^\varepsilon(y) \big)^* 
 \::\: S_y\scrM \rightarrow S_x\scrM \:.
 \eeq
In the continuum limit, one studies the EL equations expressed in terms of~$P^\varepsilon(x,y)$
asymptotically for small~$\varepsilon>0$. In this asymptotic regime, it suffices to analyze
the behavior of~$P^\varepsilon(x,y)$ near the light cone (i.e.\ if~$x$ and~$y$ have nearly lightlike separation).
An explicit analysis is possible based on the method of integration along characteristics
(see the {\em{regularized light-cone expansion}} in~\cite[Section~2.2]{cfs}
or, more generally, the {\em{regularized Hadamard expansion}}~\cite{reghadamard}).
Evaluating the resulting expressions in the EL equations~\eqref{EL},
one gets equations involving the gravitational field and the Dirac wave functions
(see~\cite[Section~4.5]{cfs}).
In models containing neutrinos, these equations give rise to the Einstein equations~\eqref{einstein}
(see~\cite[Section~4.9]{cfs}).

Before explaining how to go beyond classical gravity,
we point out that the above constructions only give a {\em{special class of examples}} of causal fermion systems
describing classical spacetimes. In general, the measure~$\rho$ defined by~\eqref{rhoeps}
will {\em{not}} be a minimizer of the causal action principle. We only know that it is an {\em{approximate}} minimizer
if the Einstein equations are satisfied.
A good strategy for getting an {\em{exact}} minimizer is to take the causal fermion system~$(\H, \F, \rho)$ as
constructed in Section~\ref{seccfslorentz} from a Lorentzian spacetime as the starting point,
and to minimize the causal action principle further by varying the wave evaluation operator~\eqref{weo}
of this causal fermion system. The resulting physical wave functions will no longer satisfy
the Dirac equation. Even more, in general a Lorentzian metric ceases to exist.
In other words, we are leaving the realm of classical gravity and differential geometry. The geometry
of the resulting {\em{quantum spacetime}} has been explored in~\cite{lqg}. We also refer the reader
interested in this direction of research to the survey articles~\cite[Section~4]{nrstg} or~\cite[Section~4]{grossmann}.

A more modest approach and more computational approach
for going beyond classical gravity is to work out {\em{small corrections}}
to the Dirac and Einstein equations described by the EL equations of the causal action principle.
This strategy was first used in~\cite{baryogenesis}, where modifications of the Dirac equations
were derived from the EL equations and the corresponding conservation laws,
giving rise to a mechanism of {\em{baryogenesis}}.
In the present paper, we proceed in the same spirit to again derive modifications of the Dirac equation,
but now giving a possible explanation for the dark universe.
We note that a systematic study of all corrections to the Dirac and Einstein equations
(also including dynamical coupling constants as envisioned in the preliminary preprint~\cite{dgc})
is still lacking and seems an interesting field of future research.

\section{The Minkowski Vacuum as a Minimizer of the Causal Action} \label{secmink}
We now specify the construction of Section~\ref{seccfslorentz} in order to describe the
Minkowski vacuum by a causal fermion system~$(\H, \F, \rho)$.
In Minkowski space, the scalar product~\eqref{print}
simplifies to the usual spatial integral over the polarized probability density,
\beq \label{printMink}
(\psi | \phi)_m = \int_{\R^3} \psi(t,\vec{x})^\dagger \phi(t,\vec{x})\: d^3x
\eeq
(where the dagger denotes the transposed and complex conjugated spinor).
Moreover, we choose the subspace~$\H \subset \H_m$ as the space spanned by all solutions
of negative frequency. In this way, we implement the Dirac sea picture.
This leaves us with the freedom to choose the regularization operator~${\mathfrak{R}}_\varepsilon$
in~\eqref{Reps}. The general strategy is to choose the regularization operator in such a way
that the resulting causal fermion system becomes a minimizer of the causal action principle.
At present, it is largely unknown how the resulting ``optimal regularization'' should look like.
A mathematically concise setting in which this question can be studied is the homogeneous
causal action principle as analyzed in~\cite{elhom}. Here we shall not enter the results of this analysis.
Instead, we merely {\em{assume}} that, by choosing a suitable regularization, we obtain an
{\em{exact minimizer}} of the causal action principle. 

This assumption has far-reaching consequences
which are crucial for the arguments in our paper. Indeed, having an exact minimizer implies that the EL
equations are satisfied without error terms.
Taking the continuum limit, it follows in particular that the the Minkowski vacuum satisfies the
Einstein equations. This means that, {\em{in the Minkowski vacuum}}, there is {\em{no dark matter}},
and the {\em{cosmological constant vanishes}}.

Taking these assumptions as our starting point, we shall compute how the situation changes
when Minkowski space is replaced by a time-periodic universe of a FLRW universe with 
a correlated initial and end state. For these computation, it is irrelevant how the
regularization of Minkowski space which realizes the minimum of the causal action principle
actually looks like. Instead, it will suffice to work with simplest possible regularization obtained by
by the replacement~$t \rightarrow t- i \varepsilon$. For this so-called
$i \varepsilon$-regularization, the kernel of the fermionic projector takes the form~\eqref{Peps}
(for more details on the $i \varepsilon$-regularization see~\cite[\S1.2.5 and~\S2.4.1]{cfs}).

\section{Dark Matter and Dark Energy in a Time-Periodic Universe} \label{secdarkmatter}
We now enter the detailed computations in flat spacetimes.

\subsection{Smearing in the Mass Parameter} \label{secsmear}
We set
\beq \label{Pmdef}
P^\varepsilon_m(p) := (\slashed{p}+m)\: \delta \big( p^2-m^2 \big)\: \Theta(-p^0)\:e^{\varepsilon p^0} \:.
\eeq
We now ``smear out'' the mass by setting
\beq \label{Psmear}
P^\varepsilon = \frac{1}{\delta m(\vec{k})} \int_{m_0+\delta n(\vec{k})}^{m_0+\delta m(\vec{k}) + \delta n(\vec{k})} P^\varepsilon_m\: dm \:,
\eeq
where~$p=(\omega, \vec{k})$.
We choose~$\delta m$ such that the width in~$\omega$ is given by
\beq \label{delomega}
\delta \omega = \frac{2 \pi}{T} \:,
\eeq
where~$T$ is the lifetime of our universe. Taking the difference of the dispersion relations
\[ (\omega - \delta \omega)^2 - |\vec{k}|^2 = (m_0+\delta m+\delta n)^2 \qquad \text{and} \qquad
\omega^2 - |\vec{k}|^2 = (m_0+\delta n)^2 \:, \]
we obtain for any momentum~$\vec{k}$ the condition
\[ (\omega - \delta \omega)^2 - \omega^2 = (m_0+\delta m+\delta n)^2 - (m_0+\delta n)^2 \:. \]
Expanding linearly, substituting~\eqref{delomega} and solving for~$\delta m$ gives
\beq \label{deltamrel}
\delta m = \frac{|\omega|}{m_0}\: \frac{2 \pi}{T} + \O \big( (m_0 T)^{-2} \big) \:.
\eeq
The function~$\delta n(\vec{k})$ gives us the freedom to ``slightly deform'' the mass shell;
this will be explained in Remark~\ref{remn} below.

In order to determine the resulting contribution to the Einstein equations
as obtained in the continuum limit, it suffices to compute the vector component of~$P^\varepsilon$.
In view of~\eqref{Pmdef} and~\eqref{Psmear}, we may pull out the Dirac matrices,
\beq \label{PT}
P^\varepsilon(p) = \slashed{p}\: R(p)\: \Theta(-p^0) \:e^{\varepsilon p^0} + (\text{scalar component})
\eeq
with
\beq \label{scrTdef}
R(p) := \frac{1}{\delta m(\vec{k})} \int_{m_0+\delta n(\vec{k})}^{m_0+\delta m(\vec{k}) + \delta n(\vec{k})} \delta \big( p^2-m^2 \big)\: dm
\eeq
(here by ``scalar component'' we mean a multiple of the identity matrix).
We next expand the $\delta$ distribution in the integrand in the mass parameter,
\begin{align*}
\delta \big( p^2-m^2 \big) &= \delta \big( p^2-m_0^2 \big)
+ \delta' \big( p^2-m_0^2 \big)\: (m^2-m_0^2) \\
&\quad\; + \frac{1}{2}\: \delta''\big( p^2-m_0^2 \big)\:(m^2-m_0^2)^2 + \O \big( (m^2-m_0^2)^3 \big) \:.
\end{align*}
Now we can carry out the integral in~\eqref{scrTdef},
\begin{align*}
R
&= \delta \big( p^2-m_0^2 \big) + \delta' \big( p^2-m_0^2 \big)\:
\frac{1}{\delta m(\vec{k})} \int_{m_0+\delta n(\vec{k})}^{m_0+\delta m(\vec{k}) + \delta n(\vec{k})}
\big(m^2-m_0^2 \big)\: dm \\
&\quad\: + \frac{1}{2}\: \delta''\big( p^2-m_0^2 \big)\: 
\frac{1}{\delta m(\vec{k})} \int_{m_0 + \delta n(\vec{k})}^{m_0+\delta m(\vec{k}) + \delta n(\vec{k})}
\big( m^2-m_0^2 \big)^2\: dm + \O\Big( \big(\delta m + |\delta n| \big)^3 \Big) \\
&= \delta \big( p^2-m_0^2 \big) + \delta' \big( p^2-m_0^2 \big)\:
\bigg( m_0 \,\big( \delta m+2 \delta n \big) + \frac{1}{3}\, (\delta m)^{2}+ \delta m\: \delta n
+ (\delta n)^2 \bigg)  \\
&\quad\: + \frac{1}{2}\: \delta''\big( p^2-m_0^2 \big)\: \bigg( \frac{4}{3}\: m_0^2 \:\Big( \delta m^2 + 3 (\delta m) (\delta n) + 3 (\delta n)^2 \Big) \bigg) + \O\Big( \big(\delta m + |\delta n| \big)^3 \Big) \:.
\end{align*}
According to~\eqref{PT}, the vector component of~$P$ is obtained
from this formula by multiplying with~$\slashed{p}$.

We next want to choose~$\delta n$ in such a way that the leading contributions~$\sim \delta'$
vanish. To this end, we choose
\beq \label{delnchoice}
\delta n = -\frac{\delta m}{2} - \frac{1}{24}\: \frac{(\delta m)^2}{m_0} \:,
\eeq
because then
\begin{align*}
m_0 \,\big( \delta m+2 \delta n \big) &= -\frac{1}{12}\: (\delta m)^2 \\
\frac{1}{3}\, (\delta m)^{2}+ \delta m\: \delta n + (\delta n)^2 &=
\frac{1}{3}\, (\delta m)^{2} - \frac{1}{2}\:(\delta m)^2 + \frac{1}{4}\:(\delta m)^2 + \O\big( (\delta m)^3 \big) \\
&= \frac{1}{12}\, (\delta m)^{2} + \O\big( (\delta m)^3 \big)
\:,
\end{align*}
as desired
(this choice will be discussed in Remark~\eqref{remn} below). We thus obtain
\beq
R =\delta \big( p^2-m_0^2 \big) + \frac{1}{6}\: m_0^2\, (\delta m)^2\: \delta''\big( p^2-m_0^2 \big)
+ \O\big( (\delta m)^3 \big) \:. \label{resultdeltam}
\eeq

\begin{Remark} {\bf{(Choice of $n$)}} \label{remn} {\em{
The ansatz~\eqref{Psmear} was already motivated in the introduction. It was
also explained why~$\delta \omega$ and~$\delta m$ are to be chosen 
according to~\eqref{delomega} and~\eqref{deltamrel} (see also~\eqref{disperse}
in the introduction). However, the choice of~$n$ according to~\eqref{delnchoice} still requires
a detailed explanation. The first summand in~\eqref{delnchoice}
can be understood immediately from the fact that we do not want to change the mean value of the mass.
More precisely, it gives rise to a smearing
\[ \frac{1}{\delta m(\vec{k})} \int_{m_0 - \frac{\delta m(\vec{k})}{2}}^{m_0 + \frac{\delta m(\vec{k})}{2}} P_m\: dm 
+ \O(\Delta m) \]
which is symmetric about~$m_0$. This seems the natural choice.

The quadratic term in~\eqref{delnchoice}
\[ -\frac{1}{24}\: \frac{(\delta m)^2}{m_0} \]
is less obvious. This choice can be understood from the requirement that the regularized
kernel~$P^\varepsilon(x,y)$ must be a minimizer of the causal action principle.
The argument goes as follows:
The corresponding contribution to the smeared $\delta$-distribution is given by
(here and in what follows, the symbol $\asymp$ denotes a selected contribution to the
expression on the left)
\begin{align*}
R &\asymp \delta' \big( p^2-m_0^2 \big)\: 2 \,m_0\, \delta n + \O\Big( \big(\delta m + |\delta n| \big)^3 \Big) \\
&= -\frac{1}{12}\: \delta' \big( p^2-m_0^2 \big)\: (\delta m)^2 + \O\Big( \big(\delta m + |\delta n| \big)^3 \Big) \:.
\end{align*}
Using~\eqref{deltamrel}, we can write this contribution as
\[ R \asymp -\frac{\pi^2}{3}\: \frac{1}{T^2}\: \delta' \big( p^2-m_0^2 \big)\: \frac{\omega^2}{m_0^2}
= -\frac{\pi^2}{6}\: \frac{1}{m_0^2 T^2}\: \omega\: \frac{\partial}{\partial \omega} \delta\big( p^2-m_0^2 \big) \:. \]
Taking the Fourier transform, the factor~$\omega \partial_\omega$
becomes the operator~$-(1+ t \partial_t)$, because
\begin{align*}
\int_{\R^4} & \frac{d^4p}{(2 \pi)^4} \: 
\omega\: \Big( \frac{\partial}{\partial \omega} \delta\big( p^2-m_0^2 \big) \Big)\: \Theta(-p^0)\: e^{ip \xi} \\
&= -i \frac{\partial}{\partial t} \int_{\R^4} \frac{d^4p}{(2 \pi)^4} \: 
\Big( \frac{\partial}{\partial \omega} \delta\big( p^2-m_0^2 \big) \Big)\: \Theta(-p^0)\: e^{ip \xi} \\
&= i \frac{\partial}{\partial t} \int_{\R^4} \frac{d^4p}{(2 \pi)^4} \: 
\delta\big( p^2-m_0^2 \big)\: \Theta(-p^0)\: \frac{\partial}{\partial \omega}  e^{ip \xi} \\
&= -\frac{\partial}{\partial t} \bigg( t \int_{\R^4} \frac{d^4p}{(2 \pi)^4} \: 
\delta\big( p^2-m_0^2 \big)\: \Theta(-p^0)\: e^{ip \xi} \bigg) \\
&= -\Big( 1 + t\: \frac{\partial}{\partial t} \Big) \int_{\R^4} \frac{d^4p}{(2 \pi)^4} \: 
\delta\big( p^2-m_0^2 \big)\: \Theta(-p^0)\: e^{ip \xi}
\end{align*}
(here we take into account only the leading contribution in~$\varepsilon$).
Thus the corresponding contribution to the regularized kernel is
\beq \label{npos}
P^\varepsilon_{\delta \omega}(x,y) \asymp 
\frac{\pi^2}{6}\: \frac{1}{m_0^2 T^2} \: i \Pdd_x \Big( \big( 1+t \partial_t \big) \T^\varepsilon_{m_0^2}(\xi) \Big) \:,
\eeq
where~$\T^\varepsilon_{m^2}(\xi)$ is the regularized Fourier transform of the lower mass shell,
\beq \label{Tepsdef}
\T^\varepsilon_{m^2}(\xi) = \int_{\R^4} \frac{d^4p}{(2 \pi)^4} \: \delta(p^2 - m^2)\: \Theta(-p^0)\: e^{\varepsilon p^0} e^{ip\xi} \:.
\eeq
The time derivative~$\partial_t$ in~\eqref{npos} changes the singularity structure on the light cone
by making the order of the singularity stronger. This is inhibited by the causal action principle,
because stronger singularities on the light cone give rise to extremely large contributions to
the causal action. Therefore, we can argue that minimizing the causal action
forces us to choose~$n$ according to~\eqref{delnchoice}.
}} \QEDrem
\end{Remark}

It remains to analyze the resulting formula~\eqref{resultdeltam} and to make the connection
to dark matter and dark energy. To this end, we now compute the corresponding contributions
to the kernel of the fermionic projector.
\begin{Prp} \label{prpsmear}
The ``smearing'' of the mass parameter~\eqref{Psmear} with~$\delta m$ and~$\delta n$
according to~~\eqref{deltamrel} and~\eqref{delnchoice} gives rise to corrections to the kernel of the
fermionic projector of the form
\begin{align}
\big(P^\varepsilon - P^\varepsilon_{m_0} \big)(x,y)
&= \frac{\pi^2}{6}\: \frac{1}{T^2}\: t^2\,P^\varepsilon(x,y) \label{c0} \\
&\quad\: - \frac{i \pi^2}{3}\:\frac{1}{T^2}\: \gamma^0 \:t\,\T^\varepsilon_{m_0^2}(\xi) \label{c1} \\
&\quad\: +\frac{i \pi^2}{12}\: \frac{1}{T^2}\:
\int_{-\infty}^\infty \epsilon(\tau)\: \gamma^0 \: \T^\varepsilon_{m_0^2}(\xi + \tau e_0) \: d\tau \label{c2} \\
&\quad\: -\frac{\pi^2}{12}\: \frac{1}{T^2}
\int_{-\infty}^\infty \epsilon(\tau) \:(t+\tau)\: P^\varepsilon(x,y + \tau e_0) \: d\tau \label{c3} \\
&\quad\: - (\text{\rm{scalar component}}) +\O \big( (m_0 T)^{-3} \big) \:, \notag
\end{align}
where~$\T^\varepsilon_{m_0^2}$ is the distributional kernel~\eqref{Tepsdef}
(and~$\epsilon$ is the sign function $\epsilon(\tau)=1$ for $\tau \geq 0$ and $\epsilon(\tau)=-1$ otherwise).
\end{Prp}
\Proof Applying~\eqref{deltamrel} to~\eqref{resultdeltam}, we obtain
\beq
R =\delta \big( p^2-m_0^2 \big) + \frac{2 \pi^2}{3}\: \frac{\omega^2}{T^2}\: \delta''\big( p^2-m_0^2 \big)
+\O \big( (m_0 T)^{-3} \big) \:. \label{result0}
\eeq
We rewrite the correction term in this equation as
\begin{align}
\frac{2 \pi^2}{3}\: \frac{\omega^2}{T^2}\: \delta''\big( p^2-m_0^2 \big)
&= \frac{2 \pi^2}{3}\: \frac{\omega^2}{T^2}\: \bigg( \frac{1}{2 \omega} \frac{\partial}{\partial \omega}
\Big( \frac{1}{2 \omega} \frac{\partial}{\partial \omega} \: \delta\big( p^2-m_0^2 \big) \Big) \bigg) \notag \\
&=\frac{\pi^2}{6}\: \frac{1}{T^2}\: \bigg( \frac{\partial^2}{\partial \omega^2} \delta\big( p^2-m_0^2 \big)
- \frac{1}{\omega}\:\frac{\partial}{\partial \omega} \delta\big( p^2-m_0^2 \big) \bigg) \:.
\label{result}
\end{align}
Now we can compute the Fourier transform of the resulting terms,
\begin{align}
&\int_{\R^4} \frac{d^4p}{(2 \pi)^4} \: 
\Big( \frac{\partial^2}{\partial \omega^2} \delta\big( p^2-m_0^2 \big) \Big)\: \Theta(-p^0)\: e^{ip \xi} \notag \\
&= -t^2 \int_{\R^4} \frac{d^4p}{(2 \pi)^4} \: 
\delta\big( p^2-m_0^2 \big)\: \Theta(-p^0)\: e^{ip \xi} \notag \\
&\int_{\R^4} \frac{d^4p}{(2 \pi)^4} \: \frac{1}{\omega}\:
\Big(\frac{\partial}{\partial \omega} \delta\big( p^2-m_0^2 \big) \Big)\: \Theta(-p^0)\: e^{ip \xi} \notag \\
&=-\frac{i}{2} \int_{-\infty}^\infty \epsilon(\tau) \bigg( \int_{\R^4} \frac{d^4p}{(2 \pi)^4} \:
\Big(\frac{\partial}{\partial \omega} \delta\big( p^2-m_0^2 \big) \Big)\: \Theta(-p^0)\: e^{ip \xi + i \omega \tau} 
\bigg) \: d\tau \label{tauint} \\
&=-\frac{1}{2} \int_{-\infty}^\infty \epsilon(\tau)\, (t+\tau)\: \bigg( \int_{\R^4} \frac{d^4p}{(2 \pi)^4} \:
\delta\big( p^2-m_0^2 \big)\: \Theta(-p^0)\: e^{ip \xi + i \omega \tau} 
\bigg) \: d\tau \:.
\end{align}
For clarity, we remark that in~\eqref{tauint} we used the distributional formula
\[ \frac{1}{\omega} = -\frac{i}{2}\:
\lim_{\delta \searrow 0} \int_{-\infty}^\infty \epsilon(\tau)\: e^{-\delta\, |\tau|}\: e^{i \omega \tau}\: d\tau \:. \]
After pulling the $\tau$-integral outside, we can take the limit~$\delta \searrow 0$, because
the expression inside the brackets has suitable decay properties in~$\tau$.
Using these formulas in~\eqref{PT}, in analogy to~\eqref{npos} we obtain the corrections
to~$P^\varepsilon_{\delta \omega}(x,y)$ given by
\[ \frac{\pi^2}{6}\: \frac{1}{T^2}\: i \Pdd_x \Big( t^2 \,\T^\varepsilon_{m_0^2}(\xi) \Big) 
- \frac{\pi^2}{12}\: \frac{1}{T^2}\: i \Pdd_x
\int_{-\infty}^\infty \epsilon(\tau)\, (t+\tau)\: \T^\varepsilon_{m_0^2}(\xi + \tau e_0) \: d\tau\:. \]
Carrying out the derivatives inside the integrand gives the result.
\QED

The corrections computed in this proposition give rise to dark matter and dark energy with
the scalings~\eqref{de} and~\eqref{dm}, as we now explain. The energy-momentum tensor of
a Dirac wave function~$\psi$ has the general form
\[ T_{jk} = \frac{1}{2} \re \,\Sl \psi \,|\, \big( i \gamma_j \nabla_k + i \gamma_k \nabla_j \big) \psi \Sr \]
(for a derivation see for example~\cite{ED}).
Noting that the kernel of the fermionic projector is composed of all physical wave functions
(see~\eqref{Psum}), the total energy-momentum tensor is given formally by
\beq \label{TP}
T_{jk}(x) = -\frac{1}{2} \re \Tr_{S_x \scrM} \Big( \big( i \gamma_j \nabla_k + i \gamma_k \nabla_j \big) P^\varepsilon(x,x) \Big) \:.
\eeq
The reason why this equation holds only formally is that it involves the energy-momentum
of all the sea states, which clearly must not be included in the Einstein equations.
The causal action principle takes care of this issue without the need of counter terms,
simply because the vacuum Dirac sea is a minimizer (as explained in Section~\ref{secmink}).
Consequently, the sea states do not show up in
the EL equations (more technically, for the vacuum Dirac sea configuration the positive and negative terms
in the Lagrangian~\eqref{Lagrange} cancel each other). 
However, {\em{changes}} to the vacuum Dirac sea configurations in general do {\em{not}} drop out of the
EL equations. More precisely, the term~\eqref{c0} changes the eigenvalues of the closed chain
only by a prefactor. Therefore, it does drop out of the Lagrangian.
The other terms~\eqref{c1}--\eqref{c3}, however, do not drop out, and we
may compute the corresponding contributions to the energy-momentum tensor simply
with the help of~\eqref{TP}.
Since the summands~\eqref{c1} and~\eqref{c2} involves the Dirac matrix~$\gamma^0$,
the contribute only to~$T^0_0$ and can be understood as dark matter~\eqref{dm}.
The summand~\eqref{c3}, on the other hand, also involves the spatial Dirac matrices,
thereby contributing to both dark energy~\eqref{de} and dark matter~\eqref{dm}.
In view of the factors~$1/T^2$ in~\eqref{c0}--\eqref{c3}, all these contributions have the
desired scaling behavior~\eqref{de} and~\eqref{dm}.

We finally point out that the contributions~\eqref{c0}--\eqref{c3} are singular on
the light cone. This means that, in order to compute how they contribute to the
EL equations of the causal action principle, one needs to employ the formalism of the
continuum limit (as outlined in Section~\ref{seccl}).
In this formalism, the prefactors becomes empirical so-called regularization parameters.
In particular, at present the signs of the prefactors are unknown.

\subsection{Periodizing the Spacetime} \label{secperiodize}
In this section, we compute the effect of the periodization~\eqref{periodized}.
The result of our analysis will be that the resulting contributions to the Einstein equations are much smaller than~\eqref{de} and~\eqref{dm}, so that the periodization is of no relevance for the dark universe.
Our starting point is again the smeared Dirac sea~\eqref{Psmear} and~\eqref{Pmdef}
with~$\delta m$ and~$\delta n$ as given by~\eqref{deltamrel} and~\eqref{delnchoice}.
But now we need to expand~$P(\xi)$ for large~$\xi$.
In preparation, we analyze the distribution~$\T^\varepsilon_{m^2}$ introduced in~\eqref{Tepsdef}.
The Fourier integral over the lower mass shell can be carried out explicitly using
Bessel functions to obtain (for details see~\cite[Proof of Lemma~1.2.9]{cfs})
\beq
\T^\varepsilon_{m^2}(\xi) = \frac{m^2}{(2\pi)^3} \frac{K_1 \big(m\sqrt{r^2+(\varepsilon+it)^2}\,\big)}{m\sqrt{r^2+(\varepsilon+it)^2}} \label{Tbessel} \:,
\eeq
where we again set~$\xi=y-x$ as well as~$t=\xi^0$ and~$r=|\vec{\xi}|$.
Expanding the modified Bessel function~$K_1$ for large values of its arguments~$t$ of~$r$, one finds that
(see~\cite[eq.~(10.40.2)]{DLMF})
\begin{align}
\T^\varepsilon_{m^2}(t,\vec{0}) &\simeq \frac{\sqrt{m} }{|t|^{\frac{3}{2}}} \: e^{-i m t} \Big(1+\O\big( |t|^{-1} \big) \Big)
\label{asytime} \\
\T^\varepsilon_{m^2}(0, \vec{\xi}) &= \frac{\sqrt{m} }{r^{\frac{3}{2}}} \: e^{-m r} \Big(1+\O\big( r^{-1} \big) \Big) \:.
\label{asyspace}
\end{align}
Finally, the regularized kernel of the fermionic projector~\eqref{Peps} is obtained by differentiation,
\beq \label{PepsT}
P^\varepsilon(x,y) = (i \Pdd_x + m)\: \T^\varepsilon_{m^2}(\xi) \:.
\eeq

The exponential decay in spacelike directions~\eqref{asyspace} explains why,
as already mentioned in the introduction after~\eqref{periodspace}, a periodization
in space does not lead to any measurable physical effects.
In timelike directions, however, one has only polynomial decay,
\[ P^\varepsilon(x,y)|_{\xi=(t,\vec{0})} \simeq 
\frac{m^\frac{3}{2}}{|t|^{\frac{3}{2}}} \: e^{-i m t} \Big(1+\O\big( |t|^{-1} \big) \Big) \:. \]
In order to compute the resulting effect of the periodization in time,
following the consideration after~\eqref{periodtnaive} we need to take into account
the smearing in the mass parameter. Therefore, our next task is to 
compute the effect of the smearing in the mass parameter for large times.
Our starting point is again the formula~\eqref{scrTdef} for the smeared lower mass shell.
But, instead of expanding in the mass parameter, we now rewrite the smearing by a convolution
in the variable~$\omega$. In position space, this convolution
corresponds to multiplication by a cutoff function in time, giving us the desired information
for large times.

We introduce the parameter~$\tau$ by the condition
\[ \omega^2 - m^2 = \big(\omega+\tau \big)^2 -m_0^2  \:; \]
it tells us about the shift in frequency as caused by the mass change.
Using that~$\omega$ is negative, we can express~$\tau$ and~$m$ in terms of each other by
\begin{align}
m(\tau) &= \sqrt{ m_0^2 + \omega^2 - \big(\omega+\tau \big)^2} = \sqrt{ m_0^2 - 2 \omega \tau - \tau^2} \\
\tau(m) &= -\omega - \sqrt{\omega^2 - m^2 + m_0^2} \:. \label{taumrel}
\end{align}
Moreover,
\[ m \: dm = (-\omega-\tau)\: d\tau \:,\qquad
dm = -\frac{\omega+\tau}{m(\tau)}\: d\tau \:. \]
This makes it possible to rewrite~\eqref{scrTdef} as
\beq \label{Texpand}
R = \frac{1}{\delta m(\vec{k})} \int_{\tau(m_0+\delta n)}^{\tau(m_0+\delta m + \delta n)} \delta \big( (-\omega+\tau)^2-|\vec{k}|^2 -m_0^2 \big)\: \Big( - \frac{\omega+\tau}{m(\tau)} \Big) \: d\tau \:.
\eeq
A direct computation using~\eqref{taumrel}, \eqref{deltamrel} and~\eqref{delnchoice} yields
\begin{align*}
\tau(m_0+\delta n) &= -\frac{\pi}{T} + m_0\, \O \Big( (m_0 T)^{-2} \big) \\
\tau(m_0+\delta m + \delta n) &= \frac{\pi}{T} + m_0\,\O \big( (m_0 T)^{-2} \big)
\end{align*}
(apart from the error terms, these relations are also obvious from~\eqref{delomega} and~\eqref{delnchoice}).
Again using~\eqref{deltamrel}, we obtain
\begin{align}
R &= \frac{m_0}{|\omega|}\: \frac{T} {2 \pi}
\int_{-\frac{\pi}{T}}^{\frac{\pi}{T}} \delta \big( (-\omega+\tau)^2-|\vec{k}|^2 -m_0^2 \big)\: \Big( - \frac{\omega+\tau}{m(\tau)} \Big) \: d\tau \; \Big( 1 + \O\big( (m_0 T)^{-1} \big) \Big) \notag \\
&= \frac{T} {2 \pi}
\int_{-\frac{\pi}{T}}^{\frac{\pi}{T}} \delta \big( (-\omega+\tau)^2-|\vec{k}|^2 -m_0^2 \big)\: d\tau \; \Big( 1 + \O\big( (m_0 T)^{-1} \big) \Big) \notag \\
&= \int_{-\infty}^{\infty} \delta \big( (-\omega+\tau)^2-|\vec{k}|^2 -m_0^2 \big)\: \hat{\eta}_T(\tau) \:d\tau
+ \O\big( (m_0 T)^{-1} \big) \:, \label{Terror}
\end{align}
where~$\hat{\eta}_T$ is a multiple times the characteristic function,
\[ \hat{\eta}_T := \frac{T} {2 \pi} \: \chi_{\big[-\frac{\pi}{T}, \frac{\pi}{T} \big]} \:. \]
In other words, the leading effect of the mass smearing for large~$m_0 T$ can be described
simply by taking the convolution in~$\omega$ with the function~$\hat{\eta}_T$.
In position space, this simply corresponds to multiplication by the Fourier transform of~$\hat{\eta}_T$ given by
\[ \eta_T(t) = \int_{-\infty}^\infty \hat{\eta}_T(\omega)\: e^{-i \omega t}\: d\omega
= \frac{T} {\pi t}\: \sin \Big( \frac{\pi t}{T} \Big) \:. \]
Therefore, the leading effect of the mass smearing for large times amounts to multiplying the kernel by
a function which is $T$-periodic and vanishes at~$t=n T$ with~$n \neq 0$.

Having clarified the effect of the mass smearing in position space for large times, we can now
perform the periodization in time. Using~\eqref{PepsT}, similar to~\eqref{periodtnaive} we consider the series
\begin{align*}
& \sum_{\tau \in T \Z} (i \Pdd_x + m)\: \Big( \eta_T(\tau)\: \T^\varepsilon_{m^2} \big( x,y + (\tau,0) \big) \Big) \:.
\end{align*}
In order to determine the scaling behavior, it suffices to consider the summand with~$\tau = T$.
Since~$\eta_T(t)$ vanishes at~$t=T$, the leading contribution scales like
\begin{align*}
&(i \Pdd_x + m)\: \Big( \eta_T(T + \xi^0 )\: \T^\varepsilon_{m^2} \big( x,y + (T,0) \big) \Big) \\
&\sim \gamma^0 \,\big( i \eta'_T(T+\xi^0) + m\: \eta(T+\xi^0)  \big) \: \T^\varepsilon_{m^2} \big( x,y + (T,0) \big) \\
& \sim \gamma^0 \: \frac{\sqrt{m} }{T^{\frac{3}{2}}} \Big( \frac{1}{T} + \frac{i m t}{T} \Big)
= \gamma^0 \: \frac{\sqrt{m} }{T^{\frac{5}{2}}} \big( 1 +  i m t \big) \:,
\end{align*}
where in the last line we used~\eqref{asytime}.
Compared to the contributions computed in Proposition~\ref{prpsmear}, these contributions
are by a scaling factor~$(m_0 T)^{-\frac{1}{2}}$ smaller and can therefore be neglected.

By direct computation, one finds that the contributions by the error terms in~\eqref{Terror} are even
smaller and can also be neglected. In order to explain how this comes about, we consider
exemplarily the contributions by the quadratic term in~\eqref{delnchoice}.
Expanding~\eqref{Texpand}, the resulting contribution to~$R$ takes the form
\[ - \frac{1}{24}\: \frac{\delta m(\vec{k})}{m_0} \int_{\tau(m_0 - \delta m/2)}^{\tau(m_0+\delta m/2)} 
\frac{\partial}{\partial \omega} \delta \big( (-\omega+\tau)^2-|\vec{k}|^2 -m_0^2 \big)\: \Big( - \frac{\omega+\tau}{m(\tau)} \Big) \: d\tau \:. \]
As explained after~\eqref{Terror}, the $\tau$-integration gives rise to the multiplication by~$\eta_T(t)$.
Compared to~\eqref{Terror}, we now have the additional factor
\[ \frac{(\delta m)^2}{m_0} \frac{\partial}{\partial \omega}
\overset{\eqref{deltamrel}}{\simeq} \frac{\omega^2}{m_0^3\, T^2} \:\frac{\partial}{\partial \omega} \:. \]
Taking the Fourier transform similar to the computation after~\eqref{result},
we obtain an additional scaling factor which is at least as small as~$(m_0 T)^{-1}$,
justifying that this contribution is negligible. For the other contributions one can argue similarly.

There is one point in the periodization which requires closer attention:
If we periodize both in space and time (as shown in Figure~\ref{figperiodize})
we also get contributions when the vector~$(\tau, \vec{\ell})$ in~\eqref{periodized}
is close to the light cone. The following consideration shows that the resulting contributions
to the kernel of the fermionic projector are also negligible:
The scaling behavior can be determined by considering the simple $i \varepsilon$ regularization
obtained by the replacement~$t \rightarrow t- i \varepsilon$. Then the Bessel function in~\eqref{Tbessel}
becomes (for details see~\cite[\S1.2.5]{cfs} or~\cite[Section~5.1 and Appendix~B]{lagrange-hoelder})
\[ K_1 \big(m\sqrt{r^2+(\varepsilon+it)^2}\,\big) \:. \]
Evaluating this function on the light cone~$t=\pm r$, one obtains exponential decay
\[ K_1 \big(m\sqrt{\varepsilon^2+2 i \varepsilon t}\,\big) \lesssim K_1 \big( m \sqrt{\varepsilon \,|t|} \big) 
\lesssim e^{- m \sqrt{\varepsilon \,|t|}} \:. \]
As a consequence, the functions~$\T^\varepsilon_{m^2}$ and~$P^\varepsilon$ decay exponentially
in lightlike directions as
\beq \label{eexp}
e^{- m \sqrt{\varepsilon T}} \:.
\eeq
This decay is very slow because the regularization length~$\varepsilon$ comes into play.
Nevertheless, we get exponential decay on cosmological scales, as is verified as follows.
Choosing~$m$ as the mass of the $\tau$-lepton, $\varepsilon$ as the Planck length and~$T$
according to~\eqref{Texperiment}, we obtain
\[ m \varepsilon \: \sqrt{m T} \approx 10 \]
and thus
\[ m^2\, \varepsilon T \approx 10\, \sqrt{mT} \approx 4 \times 10^{21} \:. \]
As a consequence, the exponential factor~\eqref{eexp} is extremely small and may be neglected.
With this in mind, in the periodization we may disregard null directions.
A more detailed estimate of the kernel of the fermionic projector in a strip around the
light cone can be found in~\cite[Appendix~B]{lagrange-hoelder}.

\section{Dark Matter and Dark Energy in an FLRW Universe} \label{secflrw}
In the previous section we explained how smearing in the mass and periodizing flat Minkowski space
gives rise to contributions to the Einstein equations which can be interpreted as
dark matter and dark energy.
We now explain how these findings can be extended to the more general and more realistic situation
of a Friedmann-Lema{\^i}tre-Robertson-Walker (FRLW) universe.
It turns out that, based on the formulation of the Dirac equation
in FRLW spacetimes in~\cite{moritz}, this generalization is rather straightforward.
Following the conventions in this paper, we choose
conformal coordinates and write the line element of a spatially homogeneous and
isotropic space-time as
\begin{equation} \label{lineelement}
ds^2 = S(\tau)^2 \Big( d\tau^2 - d\chi^2 - f(\chi)^2(d\vartheta^2 + \sin^2{\vartheta} \:d\varphi^2) \Big) .
\end{equation}
Here $\tau$ is a time
coordinate, $\varphi \in [0, 2 \pi)$ and $\vartheta \in (0, \pi)$ are angular coordinates,
and~$\chi$ is a radial coordinate.
In the three cases of a closed, open and flat universe, the function~$f$
and the range of~$\chi$ are given respectively by
\[ \left\{ 
\begin{array}{lll}
\text{closed universe:} & f(\chi) = \sin(\chi) \:,\quad & \chi \in (0, \pi) \\
\text{open universe:} & f(\chi) = \sinh(\chi) \:,\quad & \chi >0 \\
\text{flat universe:}  & f(\chi) = \chi \:,\quad & \chi >0\:.
\end{array} \right. \]
In the Dirac equation, the spatial dependence can be separated, giving rise to a spectral parameter~$\lambda$.
The time dependence is described by the ordinary differential equation
\beq \label{timeODE}
\left[ i \partial_\tau - S m \left( \!\! \begin{array}{cc} 1 & 0 \\ 0 & -1 \end{array} \!\! \right)
+ \lambda \left( \!\! \begin{array}{cc} 0 & 1 \\ 1 & 0 \end{array} \!\! \right) \right]
\left( \!\! \begin{array}{c} h_1 \\ h_2 \end{array} \!\! \right) = 0 \:.
\eeq
Our mechanism for dark matter and dark energy is driven by the solutions
whose kinetic energy is much larger than the rest mass. Therefore, the solutions can be
described by the WKB approximation (for details see~\cite[Section~3]{moritz})
\beq \label{WKB}
\begin{pmatrix} h_1 \\ h_2 \end{pmatrix}\!(\tau) = U(\tau)^{-1}
\begin{pmatrix} \displaystyle c_1 \exp \left(-i \int^\tau \sqrt{m^2 S^2 +\lambda^2} \:d\tau \right) \\
\displaystyle  c_2 \exp \left(i \int^\tau \sqrt{m^2 S^2 +\lambda^2} \:d\tau \right) \end{pmatrix} ,
\eeq
where~$U(\tau)$ is a unitary matrix which diagonalizes the matrix potential in~\eqref{timeODE}, i.e.\
\beq \label{diagonal}
U \begin{pmatrix} Sm & -\lambda \\ -\lambda & -Sm \end{pmatrix} U^{-1} =
\sqrt{m^2 S^2 +\lambda^2} \:\begin{pmatrix} 1 & 0 \\ 0 & -1 \end{pmatrix} .
\eeq

Next, it is convenient to transform from cosmological time~$\tau$ to observer time~$t$ defined by
\[ dt = S(\tau)\: d\tau \:. \]
Moreover, writing the phase in~\eqref{WKB} as~$\omega(t)\: dt$ we obtain the simple dispersion relation
\[ \omega(t)^2 - \frac{\lambda^2}{S(t)^2} = m^2 \:. \]
Therefore, we can interpret~$\lambda/S$ as spatial momentum. The only difference compared
to the dispersion relation in Minkowski space is that the momentum is not a separation constant,
but that it depends on the scale function. This also leads to a scale dependence of the frequency,
\[ \omega = \omega(t,\lambda, m) := -\sqrt{\frac{\lambda^2}{S(t)^2} + m^2} \:. \]

Locally, the kernel of the fermionic
projector in the FRLW spacetime can be written in analogy to~\eqref{Peps} as
\beq \label{PFRLW}
P_m^\varepsilon(x,y) = \int_{-\infty}^\infty \frac{d^4p}{(2 \pi)^4} \: (\slashed{p}+m)\:
\delta\Big( \omega(t)^2 - \frac{\lambda(\vec{p}\,)^2}{S(t)^2} - m^2 \Big)\: \Theta(-p^0)\: e^{\varepsilon p^0} e^{-ip (x-y)}
\eeq
(for details see~\cite[Section~5]{moritz}). Therefore, setting
\[ p^2 := \omega(t)^2 - \frac{\lambda^2}{S(t)^2} \:, \]
the mass smearing can be used exactly as described in Section~\ref{secsmear}.
The only change is that the formula for~$\delta m$
(see~\eqref{delomega} and~\eqref{deltamrel}) needs to be modified to
\[ 2 \pi = \int_0^T \Big( \omega \big( t,\lambda, m_0+\delta n(\lambda)+\delta m(\lambda) \big)
- \omega \big( t,\lambda, m_0+\delta n(\lambda) \big) \Big) \: dt \]
and thus
\[ \delta m(\lambda) = \frac{2 \pi}{\displaystyle \int_0^T \frac{m_0}{\sqrt{\frac{\lambda^2}{S(t)^2} + m_0^2}} \:dt}
+ \O \big( (m_0 T)^{-2} \big) \:. \]
Here~$t=0$ is the observer time of the big bang. The parameter~$T$ is the 
time from the big bang until it reaches its end state.
For simplicity of presentation, we usually refer to $T$ as the ``life time of our universe,'' although, strictly
speaking, it could also be smaller. This would mean that from time~$T$ on, the state of the universe
remains essentially the same. If the energy is much larger than the mass, this formula simplifies to
\beq \label{smearFRW}
m_0 \,\delta m(\lambda) = \frac{2 \pi}{\displaystyle \int_0^T \frac{S(\lambda)}{\lambda} \:dt}
+ \O \big( (m_0 T)^{-2} \big) \:.
\eeq
The resulting contributions to dark matter and dark energy
can be computed locally in a given reference frame simply by using the formula
for~$\delta m(\lambda)$ in~\eqref{resultdeltam} and computing the
corresponding contributions to the energy-momentum tensor exactly as explained at the end
of Section~\ref{secsmear}. From~\eqref{resultdeltam} we draw the following conclusions:
\bitem
\item[(1)] Since the contributions to the energy-momentum tensor as computed in Proposition~\ref{prpsmear}
are a consequence of the mass smearing, which is performed globally in spacetime,
dark matter and dark energy are evenly distributed in spacetime. Thus they do not depend on the scale
factor~$S(\tau)$.
\item[(2)] Via~\eqref{smearFRW}, the overall magnitude of the effect does depend on the global geometry.
\eitem

\section{Outlook} \label{secoutlook}
In this paper we saw that, arranging a universe with a
correlated quantum state beginning and end state by slightly modifying the frequencies of the
physical wave functions, one gets contributions to the EL equations of the causal action principle
which can be interpreted as dark matter and dark energy with the scalings~\eqref{de} and~\eqref{dm}.
Our method for modifying the frequencies was to smear out the mass parameter in the Dirac
equation and to select the frequencies compatible with the boundary conditions at the initial and final time.
It is an open question  whether our ``smearing'' of the mass parameter can be understood
in analogy to a {\em{generalized Uncertainty Principle}} like for example the Karolyhazy Uncertainty
Relation~\cite{singh2}. 
We conclude by pointing out that, although this method seems a natural and canonical starting point,
we expect that, ultimately, additional effects need to be taken into account.
For instance, other modifications of the Dirac equation were derived in~\cite{baryogenesis},
giving rise to a mechanism of baryogenesis. Moreover, it is conceivable that
the dynamical behavior of the regularization might give rise to changes of the masses and
coupling constants on cosmological scales (see~\cite{dgc} for some preliminary considerations in this direction)
A systematic study of all these corrections and their mutual interplay
seems an interesting project for future research.

\Thanks{{{\em{Acknowledgments:}} 
We are grateful to the ``Universit\"atsstiftung Hans Vielberth'' for generous support.
We would like to thank Edward Lulofs and the referees for helpful comments on the manuscript.
}

\providecommand{\bysame}{\leavevmode\hbox to3em{\hrulefill}\thinspace}
\providecommand{\MR}{\relax\ifhmode\unskip\space\fi MR }
\providecommand{\MRhref}[2]{%
  \href{http://www.ams.org/mathscinet-getitem?mr=#1}{#2}
}
\providecommand{\href}[2]{#2}

\end{document}